\documentclass{optica-article}

\journal{opticajournal} % for journals or Optica Open

\articletype{Research Article}

\usepackage{lineno}
\begin{document}

\title{Analysis of Photonic Circuit Losses with Machine Learning Techniques}

\author{Adrian Nugraha Utama,\authormark{1} Simon Chun Kiat Goh,\authormark{2} Li Hongyu,\authormark{2} Wang Xiangyu,\authormark{2} Zhou Yanyan,\authormark{2} Victor Leong\authormark{1,*} and Manas Mukherjee\authormark{1,3}}

\address{\authormark{1}National Quantum Federated Foundry (NQFF), Institute of Materials Research and Engineering (IMRE), Agency for Science, Technology and Research (A*STAR), 2 Fusionopolis Way, Innovis \#08-03, Singapore 138634, Republic of Singapore\\
\authormark{2}Institute of Microelectronics (IME), Agency for Science, Technology and Research (A*STAR), 2 Fusionopolis Way, Innovis \#08-02, Singapore 138634, Republic of Singapore\\
\authormark{3}Centre for Quantum Technologies, National University of Singapore, 3 Science Drive 2, Singapore 117543, Republic of Singapore\\
}

\email{\authormark{*}victor\_leong@imre.a-star.edu.sg} %% email address is required; see note below about the corresponding author designation

% use {asbstract*} to suppress the copyright line. Copyright information will be added in production

\begin{abstract*} 
Low-loss waveguides enable efficient light delivery in photonic circuits, which are essential for high-speed optical communications and scalable implementations of photonic quantum technologies.
We study the effects of several fabrication and experimental parameters on the waveguide losses of a silicon nitride integrated photonics platform using various machine learning techniques.
Compared to more complex machine learning algorithms, our results show that a simple linear regression model with variable selection offers a lower prediction error with high interpretability. 
The high interpretability, along with our domain knowledge, led to unique identification of fabrication process dependencies on the final outcome. With these insights, we show that by improving the process flow, it is possible to improve the loss by mitigating the cause in a real experiment.

\end{abstract*}

%%%%%%%%%%%%%%%%%%%%%%%%%%  body  %%%%%%%%%%%%%%%%%%%%%%%%%%
\section{Introduction}

Photonic integrated circuits (PICs) are important for a wide range of applications as they allow for efficient and scalable manipulation of light with compact devices. 
Silicon PICs have undergone numerous technological developments since their inception; 
their compatibility with standard fabrication processes in the semiconductor industry~\cite{siew2021review}
has led to the proliferation of wafer-scale production of silicon PICs~\cite{sia2022wafer, liu2021high, ye2023foundry}.
Silicon nitride (SiN) PICs have gained prominence in recent years for their low losses across a wide range of wavelengths, 
which is necessary for many quantum photonic applications~\cite{silverstone2016silicon,xiang2022silicon} requiring ultra low loss. 

Understanding and quantifying process variations in fabrication is critical to ensure reliable performance and high manufacturing yield~\cite{xing2022capturing, chen2013process}. 
In addition, performance is influenced by a variety of parameters related to how PICs are designed and operated. 
The application area of the PIC also plays a role: 
for example, very low loss is important for photonic quantum computing~\cite{o2007optical}, while minimizing mode crosstalk
may be critical for sensing applications~\cite{yang2014nanoscale}.
Given the sheer number of parameters involved, 
finding an optimal parameter set for the best performance for all applications is a challenge. 

To improve the accuracy of performance predictions for PICs, 
the fabrication variability can be rigorously modeled using Monte Carlo methods~\cite{lu2017performance, hamdani2023modelling}. 
Machine learning~(ML) has also recently gained significant attention in tackling such multi-parameter problems, 
and has been used to optimize production masks in manufacturing settings~\cite{yadav2023advances}, wafer hotspot detection~\cite{lin2017machine}, film deposition~\cite{kim2005use,purwins2011regression,lee2011neural},  
waveguide properties~\cite{james2021evaluating, hinum2024random},
and even waveguide simulation times~\cite{alagappan2019deep}.

Like the process flow for semiconductor device manufacturing, the fabrication of PICs at the wafer scale also undergoes design, fabrication, and testing processes. 
As the full development cycle can be expensive and time-consuming, the cost of repeating a failure is prohibitively high, and
it is essential to identify the specific cause of a failure and improve it. 
Thus, the ML method not only needs to have a high predictive power, but also needs to yield practical insights leading to effective mitigation strategies.
In this work, we investigate how photonic waveguide losses are affected by various design, fabrication, and operational parameters 
using several ML techniques. 
We evaluate the advantages and limitations of each technique, and benchmark their performances. 
Based on our results, we identified the parameter contributing most significantly to the waveguide loss, 
and directed our efforts towards mitigating the associated loss mechanism.
We obtained a significant reduction in PIC power loss,  
thus demonstrating the applicability of ML techniques to the development of integrated photonic platforms in general and a simplified model in particular to our design study.

\section{Methodology}

The PICs are fabricated on a silicon nitride (SiN) platform at the Institute of Microelectronics (IME), Singapore under the National Quantum Federated Foundry (NQFF) program. 
A SiN layer is deposited on 12-inch silicon wafers with thermal oxide using a plasma-enhanced chemical vapor deposition (PECVD) technique, after which the photonic circuits are patterned using immersion lithography and covered with a top oxide cladding. 
Some of the wafers undergo chemical-mechanical polishing (CMP) of the bottom oxide and SiN layers.

Device characterization is performed on a semi-automated optical probe station. 
Lensed fibers are coupled to the waveguide inputs and outputs, which are based on inverse taper waveguide structures. 
The fiber-waveguide coupling is optimized using a peak-search algorithm. 

\subsection{Parameters for analysis}
In this work, we define the waveguide loss as the overall insertion loss, i.e. the ratio of the output optical power (coupled to the output fiber) to the optical power from the input fiber incident on the PIC. 
This can be separated into two components: 
(a) the waveguide propagation loss 
and (b) the fiber-waveguide coupling loss.
The propagation and coupling losses can be derived from the waveguide loss measurements using the waveguide cutback technique~\cite{goh2024improved}: 
the losses (in dB) of waveguides with different lengths are measured and fitted to a linear equation considering a linear increase in the loss with the length of the waveguide. The slope and intercept of the fit represent the propagation and coupling losses, respectively.

In our ML analysis, we choose the waveguide loss as the target variable, 
instead of analyzing the propagation and coupling losses independently. 
This is for two reasons: First, using only one target variable greatly simplifies the analysis.  
Second, the propagation and coupling losses are derived values estimated by fitting the measured waveguide loss, 
which therefore have added uncertainties. 
Thus, considering the waveguide loss simplifies the analysis while maintaining the validity of the results.

While there are many parameters that could affect the waveguide loss, in this study we focus on several parameters based on our domain knowledge and specific work flow as summarized in Table~\ref{tab:exp_params}. 
In terms of the waveguide geometry, the waveguides have different lengths spanning from 1.8~cm to 13~cm, and two different widths (1.0~$\mu$m and 1.5~$\mu$m).
In terms of fabrication parameters, we compare devices fabricated with and without chemical-mechanical polishing (CMP).
Finally, in terms of operating conditions,
we compare two different wavelengths (1550~nm and 1600~nm) and two different polarizations (TE and TM). 

\begin{table}[htbp]
\centering
\caption{\bf Summary of the parameters in the study}
\begin{tabular}{cccc}
\hline
         Parameters & Data type & \multicolumn{2}{c}{Values} \\
\hline
         $x_1$: Length & Numerical & \multicolumn{2}{c}{ 1.8 cm to 13 cm} \\
         $x_2$: Waveguide Width & Categorical & 1.0~$\mu$m & 1.5~$\mu$m\\
         $x_3$: Polishing & Categorical & no CMP & with CMP\\
         $x_4$: Wavelength & Categorical & 1550~nm & 1600~nm\\
         $x_5$: Polarization & Categorical & TE & TM\\
\hline
    \end{tabular}
    \label{tab:exp_params}
\end{table}

\subsection{Machine Learning Analysis and Validation}

A machine learning (ML) algorithm learns the patterns of an existing dataset (training data) that can be generalized and applied to unseen data (test data). 
In the branch of supervised ML, a target variable $y$ can be predicted from a collection of predictor variables $x_i$ (also known as features) 
by training on a large amount of labeled data (also known as the training data). 
A model with excellent predictive power enables reliable forecasting for new (or test) data and rapid data-driven decision making. However, if the features of the model are too convoluted, the correlations between features and the real-world practical variable cannot be interpreted correctly.
Therefore, models with good interpretability can be used to extract meaningful insights into how the system is influenced by the practical variables.

In this study, we utilize various supervised ML methods, ranging from simple linear regression models to complex non-linear models. 
Specifically, the task is to predict the waveguide loss $y$ from the features $x_i$, 
which are described in Table~\ref{tab:exp_params}. 
The features may be varied intentionally in the design or unintentionally as a consequence of fabrication tolerance, 
but this difference does not affect the ML outcome.
The measured data in our dataset is almost equally distributed across the different features. 
For each unique combination of features, we measure two identical devices from different locations on the same wafer;
any discrepancy in the waveguide loss between the identical devices can be solely attributed to fabrication variability. 
A small fraction ($<5\%$) of the devices are rejected from the analysis as they are considered to be defective ($>40$~dB loss). 
In total, we included 210 measured devices for our analysis. 

As indicated in Table~\ref{tab:exp_params}, we studied one numerical variable (quantitative values, continuous over its range) and four categorical variables (has a specific set of possible values). 
If we extend our study to include more values of waveguide widths or wavelengths, the associated data type could remain as categorical, or be changed to numerical. 
It is important to note that the results might differ, as numerical variables imply a continuous range between the values, while categorical variables do not. 

To characterize the performance of the machine learning models, we use the root mean square error (RMSE) between the predicted values and the actual outcomes of the waveguide loss $y$. 
As machine learning models are prone to overfitting, we use the K-fold cross-validation technique with K\,=\,10, where we divide the dataset into K equal parts, train the model using the K-1 parts, and use the training result to validate the unseen data from the remaining part~\cite{kohavi1995study}.
We repeat the cross-validation procedure to cover all K parts. The training RMSE is the average of the RMSE of the K-1 parts during training, while the validation RMSE is the average of the RMSE of the validation data. 

\section{Machine Learning Models}
\label{sec:machine_learning_models}

The supervised ML models in our study are selected for their wide usage and ease of implementation, as a representative sample of available options.
First, we present simple linear regression models, 
including variants with interaction terms to account for higher data complexity. 
We also consider variable selection and regularization to reduce model complexity and highlight the significant features.

Next, we explore some widely-used ML regression algorithms, such as decision tree, random forest, support vector, and kernel ridge regressions. 
Decision tree and random forest regressions work by exploring various different decision paths to predict a discrete outcome, and are therefore highly nonlinear. 
Support vector and kernel ridge regressions work by finding effective vector structures to model the data, and hence reduce the noise contribution.

Finally, we use a simple artificial neural network with non-linear activation functions, 
which is trained to minimize the output error via back-propagation. 
Such neural networks have proven to be very popular, especially for applications involving a massive amount of training data, but generally lack interpretability.

\subsection{Linear Regression Models}
\label{section:lr-models}

A basic model that can be used to describe the relationship between waveguide loss and the features is linear regression (LR). 
Here, the waveguide loss $y$ has a linear dependence on the different features $x_i$ as described in Table~\ref{tab:exp_params}, and can be written as:
\begin{equation}
    y = b + \sum_i b_i x_i,
\end{equation}
where $b$ is the intercept and $b_i$ are the regression coefficients. 
The optimal regression coefficients are obtained by fitting the experimental data to the model using the least squares method~\cite{james2023linear}. 
In the basic LR model, the training RMSE is 3.57~dB (coefficient of determination $R^2=0.831$), while the validation RMSE is 3.69~dB. 
This serves as our baseline result.

To capture a more complex relationships between different features, we can include the interaction terms as follows:  
\begin{equation}
    y = b + \sum_i b_i x_i + \sum_{i\neq j} b_{ij} x_i x_j + ... + b_{12345} x_1 x_2 x_3 x_4 x_5 . 
\end{equation}
For simplicity, we choose to not include self-interacting terms ($x_i^2$, $x_i^3$, etc.) as such interaction mechanisms are not expected to be present in our devices as part of our domain knowledge.  
Here, both the training and validation RMSE are reduced significantly to 0.97~dB ($R^2=0.987$) and 1.17~dB, respectively. 
This shows that LR with interaction terms is a better choice than the basic model, 
but offers limited insights beyond its predictive power. 

Another issue we observed is that most of the estimated regression coefficients are too small to contribute meaningfully to the outcome, and could potentially reduce the model accuracy~\cite{khalid2014survey}. 
This can be mitigated with the variable selection methods explored in the next two sections.

\subsection{Linear Regression with Statistical Variable Selection}
\label{section:lr-variable}

Using statistical tests in LR can  
be problematic; a common method such as stepwise regression typically results in biased estimates with overestimated significance~\cite{harrell2012regression}. 
Instead, we use the statistical measures, namely the p-value and F-statistic~\cite{james2023linear}, solely as proxies to rank the importance of the variables (features), 
rather than to directly draw any statistical conclusions. 

For each cross-validation run, we rank either the p-value or the F-statistic of the regression coefficients using the LR with interaction model as described in the previous section. We then select the top $N$ variables and reapply the linear regression with only the selected variables.
To select the best-tuned model, we vary $N$ as a hyperparameter to obtain the lowest RMSE using either the p-value and F-statistic ranking. 

\begin{figure}[ht!]
\centering\includegraphics[width=7cm]{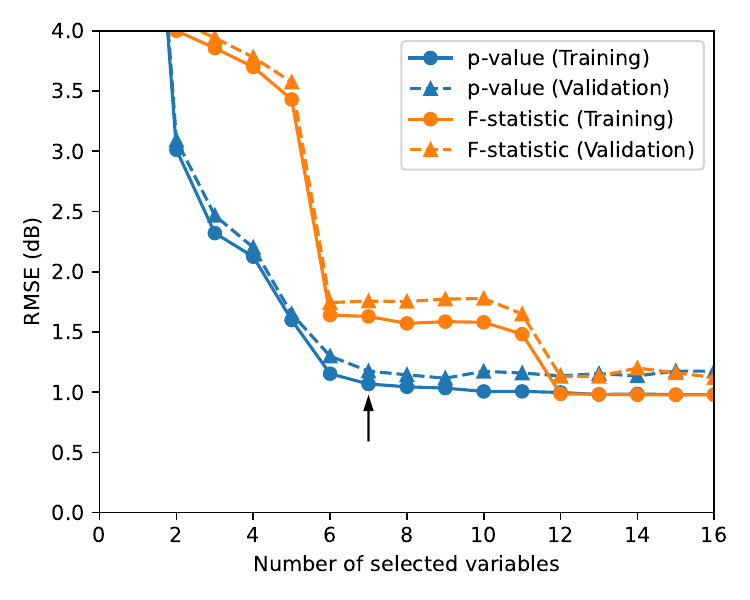}
\caption{
The performance of the LR with variable selection model 
for different numbers of selected variables $N$. 
Variable selection via p-value ranking generally outperforms F-statistic ranking. 
With larger $N$, the gap between training and validation errors widens, which indicates overfitting. 
We select $N=7$ with p-value ranking as our tuned model.}
\label{fig:lr_with_variable_selection}
\end{figure}

Figure \ref{fig:lr_with_variable_selection} shows how the RMSE varies with $N$.  
We choose the model with $N=7$ selected via p-value ranking as our tuned model.  
The training and validation RMSE are 1.07~dB and 1.14~dB, respectively. 
Although these values are close to the LR with interaction model, the LR with variable selection model here is less prone to overfitting, as the difference between training and validation errors is reduced. 

\begin{table}[tbp]
\centering
\caption{\bf Top variables ranked by the p-values, with the corresponding coefficient estimates of the tuned model ($N=7$).}
\begin{tabular}{cccc}
\hline
         Variable & Coefficient & p-value & Coef. Estimate \\
\hline
         WG Length & $b_1$ & 1e-46 & 3.44(3) dB/cm\\
         WG Length * $\lambda$ [1600~nm] & $b_{14}$ & 1e-26 & -2.34(4) dB/cm\\
         WG Length * Polarization [TM] & $b_{15}$ & 1e-6 & -1.05(2) dB/cm\\
         Intercept & $b$ & 1e-5 & 4.3(1) dB\\
         WG Length * $\lambda$ [1600~nm] * Polarization [TM] & $b_{145}$ & 1e-3 & 0.85(2) dB/cm\\
         WG Length * Polishing [with CMP] & $b_{13}$ & 0.05 & -0.30(1) dB/cm\\
         WG Width [$1.5~\mathrm{\mu m}$] & $b_{2}$ & 0.45 & 0.4(1) dB\\
\hline
WG: waveguide; $\lambda$: wavelength
    \end{tabular}
    \label{tab:variable_selection}
\end{table}

Table~\ref{tab:variable_selection} lists the most important variables ranked by their p-values, 
along with the estimated regression coefficients for the tuned and refitted model with $N=7$. 
The most important variable is the waveguide length, represented by the regression coefficient $b_1$; 
followed by coefficient $b_{14}$, which can be interpreted as a change of the propagation loss when the wavelength of 1600~nm is used
instead of 1550~nm. 
With $b_{14}=-2.34(4)$~dB/cm (compared to $b_{1}=3.44(3)$~dB/cm), the expected reduction in propagation loss is substantial with this change in wavelength.
Having thus identified the wavelength as a major variable affecting the propagation loss, 
we discuss a mitigation strategy in Section~\ref{sec:applying_result}.

With a ranking-based variable selection, this model also serves as a tool for analyzing the key features. 
Furthermore, the regression terms are straightforward to interpret, and can be
factored into terms that affect either the propagation losses or the coupling losses. 
Depending on the application, the outcome and predictor variables can also be generalized to different data types using a generalized linear model approach~\cite{nelder1972generalized}. 

\subsection{LASSO Regression}

LASSO (least absolute shrinkage and selection operator) is a regression technique that incorporates regularization to reduce the model complexity~\cite{tibshirani1996regression}. 
Here, we start with the linear regression (LR) with interaction model (see Section~\ref{section:lr-models}) and introduce a $L^1$ penalty term 
$\alpha \sum_j |b_j|$ to the loss function. 
This dampens the strengths of the coefficients and shrinks some of them to zero, effectively removing them from the model and preventing overfitting.

LASSO regression is similar to LR with variable selection (see Section~\ref{section:lr-variable}) in that both methods reduce the number of coefficients. 
However, instead of selecting the variables by a ranking method, the parameters in LASSO regression are  
selected via the strength of the regularization parameter $\alpha$. 

\begin{figure}[tb!]
\centering\includegraphics[width=7cm]{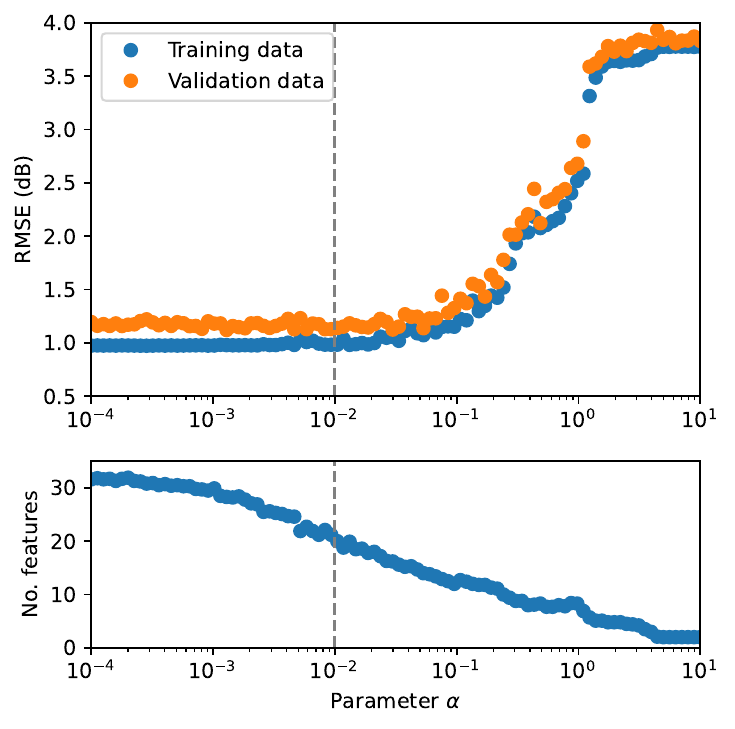}
\caption{The training and validation RMSE (top) with respect to the regularization parameter $\alpha$ with the corresponding number of selected features (bottom). With lower regularization (smaller $\alpha$), the validation error is higher, albeit slightly, than the training error, indicating overfitting. 
With higher regularization, the errors are much higher as the model becomes too simple. 
We select $\alpha = 0.01$ as our tuned model (gray dotted line), as the validation error is both low and relatively stable in its vicinity.
}
\label{fig:lasso_tuning}
\end{figure}

The tuning of $\alpha$ is shown in Figure~\ref{fig:lasso_tuning}. 
We choose $\alpha=0.01$ as the tuned model, where the training and validation RMSE are 1.02~dB and 1.16~dB, respectively. 
While the performance is similar to LR with variable selection, 
LASSO requires around 20 selected features (non-zero regression coefficients), which is higher than the 7 coefficients needed in the former case. 
This could explain the slight overfitting observed here.

\subsection{Decision Tree and Random Forest Regression}

A decision tree model 
branches a dataset into smaller and smaller subsets using feature-based conditions that minimize the overall error~\cite{quinlan1986induction}. 
The outcome is obtained by following decision paths from the root to the leaf of the decision tree. 
Since different paths are distinct, the model is inherently non-linear. 
A random forest model consists of an ensemble of decision trees, each trained with random sub-samples of the whole dataset (typically obtained via bootstrapping)~\cite{breiman2001random}. 
This allows for better model generalization and less overfitting.

We implement the decision tree and random forest regression using the widely used open-source library \textsc{scikit-learn}~\cite{pedregosa2011scikit}. 
An important hyperparameter to be tuned is the maximum tree depth, which describes the maximum number of decisions that can be made from the root to the leaf. 
Higher tree depths may correspond to more overfitting, as the model would try to distinguish increasingly minute details between the subsets. 
For the other hyperparameters, we use the default values as provided by the library. 

\begin{figure}[tb!]
\centering\includegraphics[width=7cm]{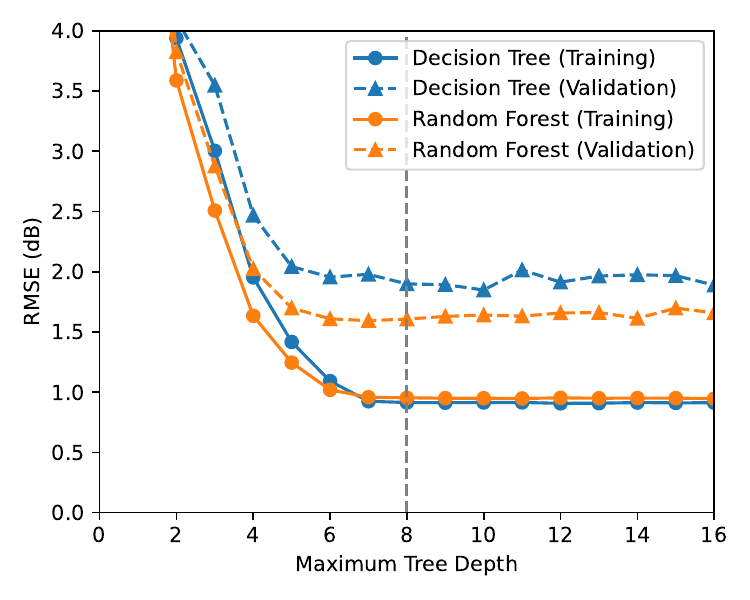}
\caption{The training and validation RMSE of the decision tree and random forest regression with respect to the maximum tree depth. 
While the training RMSE are similar for both models, random forest regression performs better during validation. 
In our data, we do not observe overfitting for trees with maximum depths beyond 10.
We select the tuned maximum tree depth of 8 (gray dotted line) for both decision tree and random forest models.}
\label{fig:tree_and_forest}
\end{figure}

The tuning of the maximum tree depth is shown in Figure~\ref{fig:tree_and_forest}. 
We select the tuned maximum tree depth of 8 for both decision tree and random forest models. 
For the tuned decision tree model, 
the training and validation RMSE are 0.92~dB and 1.89~dB, respectively. 
For the tuned random forest model, 
the training and validation RMSE are 0.95~dB and 1.66~dB, respectively. 
We can also quantify the importance of each feature in determining the outcome. 
For both models, the features ranked in order of importance are: 
waveguide length ($x_1$), wavelength ($x_4$), polarization ($x_5$), polishing ($x_3$), and waveguide width ($x_2$). 
The respective feature importance percentages are $\left[51\%, 42\%, 5.5\%, 1.6\%, 0.2\%\right]$ for the decision tree and $\left[51\%, 42\%, 5.1\%, 1.4\%, 0.3\%\right]$ for the random forest.

\subsection{Support Vector and Kernel Ridge Regression}

Support vector regression (SVR) and kernel ridge regression (KRR) aim to find a function to accurately predict the relationship between the outcome and features, 
by using kernel methods to transform the data into a high-dimensional space, which enables complex patterns to be captured.
SVR uses a set of support vectors to construct hyperplanes on which to perform regression via a kernel function~\cite{smola2004tutorial}. 
SVR allows for a deviation $\epsilon$ from the hyperplanes, where the error is not being counted. This allows the model to be less sensitive to noise. 
Besides the parameter $\epsilon$, it uses the regularization parameter $C$ to determine the model complexity.
On the other hand, KRR uses a squared error loss for all the data points instead of having error-free $\epsilon$-regions as in SVR~\cite{vovk2013kernel}. Thus, it is effectively a ridge regression combined with the kernel method, with a regularization parameter $\alpha = 1/2C$.   

Our implementation of SVR and KRR uses the \textsc{scikit-learn} library~\cite{pedregosa2011scikit}. 
We use a radial basis function kernel, as it has a more localized effect on the features and are easily adaptable to various data types and distributions. 
During the pre-processing step, we normalize the features so that they have comparable scale. For SVR, we tune both the $\epsilon$ and $C$ hyperparameters, while for KRR we only tune the $\alpha$ hyperparameter. 
For the other hyperparameters, we use the default values as provided by the library. 

\begin{figure}[tb!]
\centering\includegraphics[width=7cm]{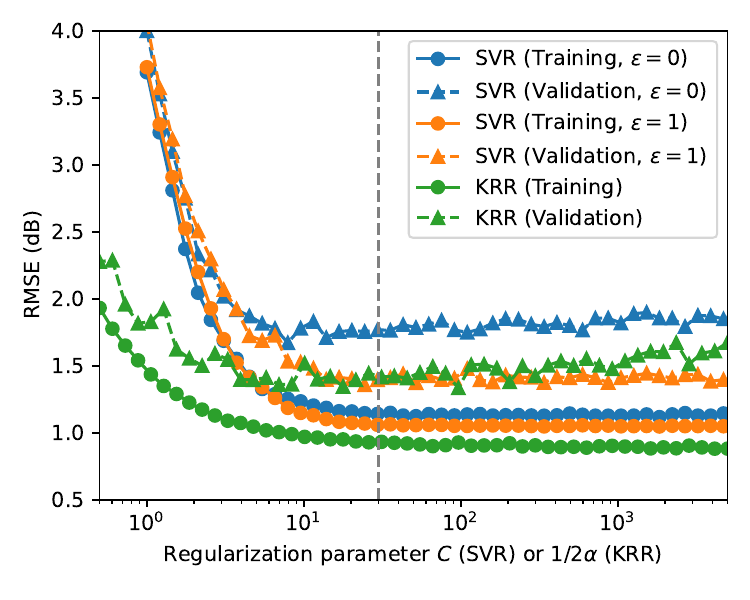}
\caption{The training and validation RMSE of the SVR and KRR models with respect to the regularization parameters, with increased regularization at lower values on the horizontal axis. 
For SVR, we vary $\epsilon$ and find that the validation curve with the lowest error is achieved at $\epsilon=1.0$ (orange lines). As a comparison, the model with no error-free regions at $\epsilon=0$ exhibits higher error (blue lines). The performance for the KRR model (green lines) is close to the SVR model, and we select $C=1/2\alpha=30$ as our tuned models.}
\label{fig:svr_krr}
\end{figure}

Figure~\ref{fig:svr_krr} depicts the tuning of the regularization parameters for the SVR and KRR models. For the tuned models, the training RMSE are 1.06~dB and 0.96~dB for the SVR and KRR models, respectively. 
The validation RMSE are 1.37~dB for both models. The tuned SVR model averages about 88 support vectors. 

\subsection{Artificial Neural Network}

An Artificial neural network (ANN) is a ML model consisting of layers of interconnected units known as neurons, as inspired by their biological counterpart. 
Each neuron processes information from the inputs and applies a non-linear function to generate its output. 
The output is multiplied by a parameter which represents the connection strength between the neurons, then used as the input for the next layer of neurons, and so on, until the last layer where the output variable is estimated. 
A fully connected system links all the neurons between the layers. 
During the training procedure, the connectivity parameters are adjusted via the back-propagation algorithm to improve the accuracy of the model.

We use \textsc{TensorFlow}~\cite{tensorflow2015-whitepaper} to implement the ANN. 
We choose a simple network architecture with a few uniform layers, each containing a set amount of neurons, corresponding to the hyperparameters n\_layers and n\_neurons. 
The neurons are fully connected between the layers. 
The network is trained by minimizing the RMSE of the waveguide loss using the Adam optimization algorithm~\cite{kingma2014adam}. 
We use the default parameters as documented in \textsc{TensorFlow} version 2.19.0, and tune the hyperparameters n\_layer and n\_neuron to obtain a tuned model. 
We note that this tuned model may not be the most optimal ANN model, as there is a vast number of model architectures and optimization algorithms that are not explored here.

\begin{figure}[ht!]
\centering\includegraphics[width=9cm]{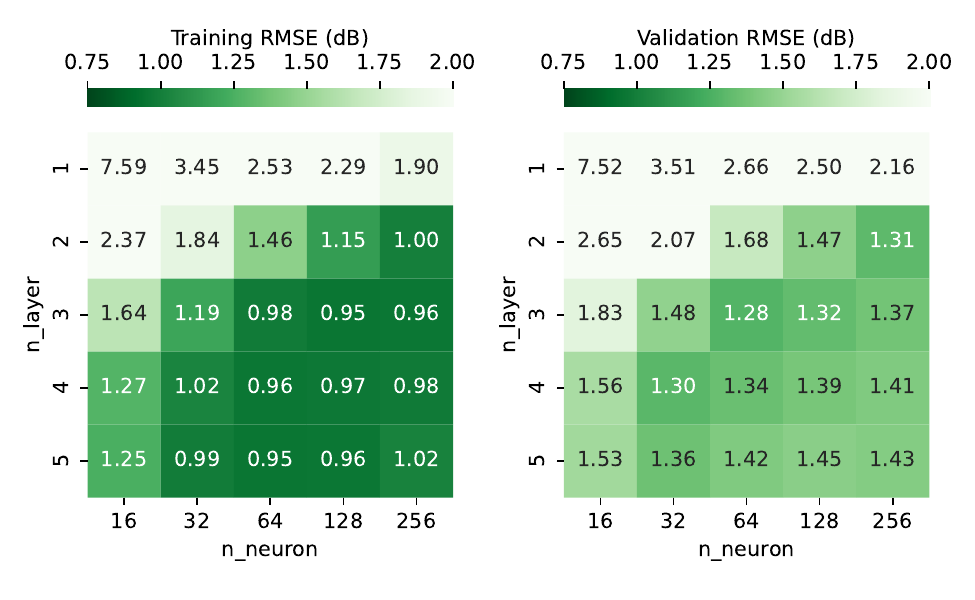}
\caption{The training (left) and validation (right) RMSE heat maps of our ANN model, with respect to the hyperparameters n\_layer and n\_neuron. 
The training RMSE shows a leveling-off behavior for relatively complex models with many layers and neurons (the dark green region), while the validation RMSE shows overfitting behavior. We choose n\_layer\,=\,3 and n\_neuron\,=\,64 as our tuned model.}
\label{fig:ann_tuning}
\end{figure}

Figure~\ref{fig:ann_tuning} shows how the number of layers and neurons affect the training and validation RMSE for our ANN model. 
For the tuned model, the training and validation RMSE are 0.98~dB and 1.28~dB, respectively.

\section{Benchmarking}
\label{sec:model_evaluation}
 
For each of the ML models above, the K-fold cross-validation technique was used to optimize the hyperparameters
to obtain tuned models with a low overall error and minimal overfitting. 
However, to compare across the different ML models, a simple K-fold cross-validation may not be adequate for confidence interval estimation, 
as it tends to underestimate the confidence intervals due to unaccounted correlations in the resampled data~\cite{bengio2003no, vanwinckelen2012estimating}. 
Here, we follow the approach outlined in Ref.~\cite{bates2024cross} and perform nested cross-validation (NCV), 
which utilizes an inner cross-validation loop with an uncorrelated outer loop to derive unbiased estimates. 
We choose the settings of 10-fold nested cross-validation with 50 repetitions. 
It is worth noting that NCV is computationally intensive due to the inner cross-validation loops and multiple repetitions. 

Table~\ref{tab:model-summary} shows the result of the NCV RMSE for the different models. 
Each NCV value is reported as a range spanning a central value, where the range signifies the 95\% confidence interval.  
Models with a smaller range 
perform more consistently, i.e. the error fluctuates less when training or predicting with different sets of data drawn from the same ensemble. 

Unlike in hypothesis testing, overlaps in the confidence intervals do not indicate that the models are not statistically different in terms of their error performance. 
For instance, while the NCV RMSE range of LASSO regression is completely overlapped by that of the decision tree regression, 
we argue that LASSO is the more suited model for our data: 
though there is a chance that decision tree regression yields lower error than LASSO for some combination of training and validation data,
LASSO performs much more consistently.

Therefore, we propose to benchmark the models using the upper bound (UB) of the NCV RMSE range;
the predictive error of the model is expected to be less than the UB error in the majority of cases ($\sim97.5\%$). 
Thus, we consider models with lower UB errors to have a better overall performance. 

The training NCV RMSE is also reported Table~\ref{tab:model-summary}. 
These values are equivalent to the training RMSE from cross-validation as described in Section~\ref{sec:machine_learning_models}, up to some small noise due to randomization. 

\begin{table}[htbp]
\centering
\caption{\bf Performance Summary of Various Machine Learning Models}
\begin{tabular}{cccc}
\hline
Models & NCV RMSE (dB)  & UB Error (dB) & Training Error (dB) \\
\hline
Linear Regression (LR) & 3.70 $\pm$ 0.52 & 4.22 & 3.57 \\
LR with Interaction & 1.17 $\pm$ 0.14 & 1.31 & 0.97 \\
LR with Variable Selection & 1.15 $\pm$ 0.18 & 1.33 & 1.07 \\
LASSO Regression & 1.14 $\pm$ 0.24 & 1.38 & 1.00 \\
Decision Tree Regression & 1.91 $\pm$ 1.44 & 3.35 & 0.91 \\
Random Forest Regression & 1.63 $\pm$ 1.06 & 2.69 & 0.95 \\
Support Vector Regression & 1.38 $\pm$ 0.57 & 1.95 & 1.06 \\
Kernel Ridge Regression & 1.36 $\pm$ 0.73 & 2.09 & 0.95 \\
Artificial Neural Network & 1.27 $\pm$ 0.58 & 1.85 & 0.98\\
\hline
\end{tabular}
NCV: nested cross-validation; UB error: upper bound of the NCV RMSE range
  \label{tab:model-summary}  
\end{table}

The model with the lowest UB error is LR with interaction, 
closely followed by LR with variable selection.
Nonetheless, given the substantial insights with clear interpretability (see Section~\ref{section:lr-variable}),
we regard LR with variable selection as the most useful ML model in this study.  

While still outperforming the baseline model, other machine learning models underperform compared to the linear regression-based models.
This can be understood given the high linearity in our dataset.
Tree-based models rely on distinct decision pathways, making them more suitable for dataset requiring sequential decision logic. 
SVR and KRR models perform well with highly nonlinear data, but are sub-optimal with the highly linear dataset here. 
Although the ANN outperforms the other nonlinear models, it still lags behind linear regression-based models. Moreover, the interpretability in such a black-box model is low.

\section{Applying our Results: Improving the Waveguide Loss}
\label{sec:applying_result}

From the results of the best performing model (LR with variable selection), we identify the wavelength as a significant parameter affecting the waveguide propagation loss. 
Specifically, the propagation loss at 1550~nm (C-band) is higher than at 1600~nm. 
Nonetheless, noting that operation at C-band wavelengths is critical to many applications, 
we apply this insight obtained from our analysis to focus on investigating the wavelength-dependent propagation loss, 
with the aim of reducing the loss at C-band wavelengths.

For SiN PICs, it is known that there is an absorption loss mechanism peaking at around 1520~nm due to N-H stretching bonds~\cite{worhoff1999plasma, ay2004comparative}, which explains the higher propagation loss at 1550~nm. 
Several studies have also shown that annealing the PICs at temperatures over 1000$^{\circ}$C could reduce the associated absorption loss for SiN waveguides deposited via LPCVD~\cite{henry1987low} and PECVD~\cite{wang2018nonlinear} techniques.

\begin{figure}[ht!]
\centering\includegraphics[width=10.5cm]{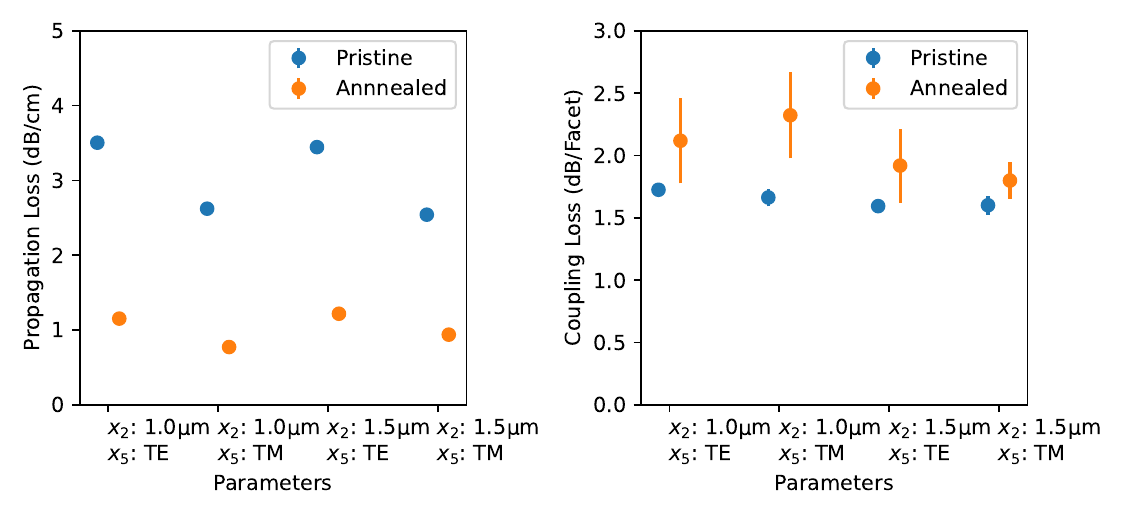}
\caption{The measured propagation loss (left) and coupling loss (right) of several pristine (blue) and annealed (orange) PICs, each corresponding to a distinct set of parameters. 
All devices here are fabricated with $x_3:$~no CMP and are measured at $x_4:$~1550~nm. 
}
\label{fig:annealing}
\end{figure}

To investigate if this approach is applicable to our PICs, 
we perform a post-fabrication annealing procedure to some of our devices by placing them inside an oven at 1050$^{\circ}$C for 1~hour.  
Our results show that the propagation loss decreases significantly after annealing (see Figure~\ref{fig:annealing}) 
by around 2~dB,
resulting in a similar performance as if the waveguides are operated at 1600~nm. 
On the other hand, the waveguide coupling loss increases slightly after annealing, along with a larger non-uniformity measured across devices, 
which might be unfavourable for precision applications. 

Other approaches to reduce the loss at C-band wavelengths include SiN deposition using a high-temperature low-pressure chemical vapor deposition (LPCVD) process~\cite{huang2014cmos} or a low-temperature NH$_3$-free PECVD process~\cite{bucio2016material}. 
With more data, we can extend our model to analyze the annealing or other procedures to enable more informed decisions towards further PIC development efforts. 

\section{Discussion}

This study yields several notable observations. 
Although the simple linear regression (LR) model gave the largest errors, 
the LR with variable selection model outperforms more complex ML models, while providing practical insights with its results.
In fact, the top three models with the lowest errors are all variants of LR models; 
the fourth-placed model is the artificial neural network (ANN), with a significantly larger error than the top three. 
While the ANN is a powerful tool for modeling complex and non-linear relationships~\cite{liu2017survey}, 
it is not necessarily the optimal choice for any multi-parameter analysis.
Although the decision tree and random forest could yield the same highly interpretable results as the LR with variable selection, the errors are much higher.
Our results indicate that for datasets with highly linear relationships between variables, 
LR-based models can be a more appropriate choice.

As datasets with a high degree of linearity are common in natural science and engineering,
it may be necessary to consider the sources of (non-)linearity when applying ML analysis. 
For instance, within our study, the waveguide loss $y$ and waveguide length $x_0$ are, under normal operating conditions, related by the Beer-Lambert law~\cite{swinehart1962beer},
assuming the other parameters ($x_1$ to $x_4$) are kept constant. 
As such, models based on LR may be better suited to datasets with an inherent linearity imposed by physical laws, 
while complex models such as ANN would have to uncover this linearity at greater computational cost and with a potentially higher error.

Our dataset currently has a limited number of features, making it easier to develop LR-based models with interacting variables. 
In a sparse dataset with many more features, the complexity may render this method infeasible
as the number of interaction terms grows exponentially with the number of features. 
In such cases, we may have to limit the number of higher-order terms or perform dimensionality reduction techniques. 
As complexity increases, even for highly linear datasets, ANN might begin to outperform LR.

In future works, it may be useful to investigate hybrid machine learning models, where physical laws (such as Beer-Lambert) are considered alongside conventional machine learning techniques. 
This could be implemented with methods akin to setting the constraints in Lagrange multipliers, or running a secondary routine to minimize the deviation from the relationships given by the physical laws.

\section{Conclusion}

We apply several machine learning (ML) techniques to study how various design, fabrication, and operational parameters affect the waveguide loss of a photonic integrated circuit. 
We identify that a linear regression (LR) model with variable selection performs better than other complex ML techniques. 
The highly interpretable results of the model also allowed us to identify the operating wavelength as an important parameter, 
and thus focus our attention to mitigate the wavelength-dependent loss via post-fabrication annealing, 
which we have successfully demonstrated. 
We attribute the performance of such a simple model to the high linearity and low dimensionality of our dataset, 
which is a common scenario in many engineering applications.

\begin{backmatter}
\bmsection{Funding}
% Content in the funding section will be generated entirely from details submitted to Prism. Authors may add placeholder text in this section to assess length, but any text added to this section will be replaced during production and will display official funder names along with any grant numbers provided. If additional details about a funder are required, they may be added to the Acknowledgment, even if this duplicates some information in the funding section. For preprint submissions, please include funder names and grant numbers in the manuscript.

We acknowledge the funding from %the Agency for Science, Technology and Research (A*STAR), Singapore
%(C230917005), and 
the National Research Foundation, Singapore, through the National Quantum Office, hosted in A*STAR, under its 
Quantum Engineering Programme (W21Qpd0307, W24Q3D0003).

\bmsection{Disclosures}
The authors declare no conflicts of interest.

\bmsection{Data Availability Statement}
Data underlying the results presented in this paper are not publicly available at this time but may be obtained from the authors upon reasonable request.

\end{backmatter}

\bibliography{sample}

\end{document}